\documentclass{TCAM}

\usepackage[english]{babel}      % para texto em Português

\usepackage[utf8]{inputenc}    % para acentuação em Português UTF-8
\usepackage{amsmath}
\usepackage{bigints}
\usepackage{epsfig}
\usepackage{multirow}
\usepackage{array}
\usepackage{url}
\usepackage{graphics}
\usepackage{subfigure}
\usepackage{psfrag}

\usepackage{framed}
\usepackage{tikz}
\title{Identification of a time-varying $SIR$ Model for Covid-19}

\runningtitle{Identification of a time-varying $SIR$ Model for Covid-19}

\author{W. M. HASELEIN	\aff{Department of Pure And Applied Mathematics, Federal University of Rio Grande do Sul, Av. Bento Gonçaves, 9500, 91509-900, Porto Alegre, RS, Brazil. E-mail: wmhaselein@gmail.com}, 
D. ECKHARD 	\aff{Department of Pure And Applied Mathematics, Federal University of Rio Grande do Sul, Av. Bento Gonçaves, 9500, 91509-900, Porto Alegre, RS, Brazil. E-mail: diegoeck@ufrgs.br} 
}

\abstracttcam{Throughout human history, epidemics have been a constant presence. Understanding their dynamics is essential to predict scenarios and make substantiated decisions. Mathematical models are powerful tools to describe an epidemic behavior. Among the most used, the compartmental ones stand out, dividing population into classes with well-defined characteristics. One of the most known is the $SIR$ model, based on a set of differential equations describing the rates of change of three categories over time. These equations take into account parameters such as the disease transmission rate and the recovery rate, which both change over time. However, classical models use constant parameters and can not describe the behavior of a disease over long periods. In this work, it is proposed a $SIR$ model with time-varying transmission rate parameter with a method to estimate this parameter based on an optimization problem, which minimizes the sum of the squares of the errors between the model and historical data.  Additionally, based on the infection rates determined by the algorithm, the model's ability to predict disease activity in future scenarios was also investigated. 
Epidemic data released by the government of the State of Rio Grande do Sul in Brazil was used to evaluate the models, where the models shown a very good forecasting ability, resulting in errors for predicting the total number of accumulated infected persons of $0.13\%$ for $7$ days ahead and $0.6\%$ for $14$ days ahead.
}
\keywords{COVID-19, Simulation, Identification, Prediction}

\acknowledgments{This work was partially supported by Coordenação de Aperfeiçoamento de Pessoal de Nível Superior - CAPES - Finance Code 001 and CNPq . Both sources belong to the Brazilian government.} 

\begin{document}
	
	\maketitle
	
	%********************************************************
	\section{Introduction}
	
	Humanity coexists with infectious diseases since the beginning of the history. Some of these infirmities can reach simultaneously a large number of individuals in the same area. When these outbreaks of contamination occur, that place is facing an epidemic. The prevalence and effects of an epidemic can diversify a lot. In less developed countries, even low risk diseases can have serious consequences. The economic development defines, in general, the quality of health cares and the efficiency to watch and contain a infectious disease. Because of that, worrying rates of mortality are still registered for malaria, typhus, cholera and others easily treatable infirmities \cite{Fred:2010}. Furthermore, an epidemic situation can affect the economy of some regions. In some cases, control measures change people circulation and the operation of commercial establishments, generating negative economic impact during the application of these measures. Thereby, understanding the phenomenon is crucial to determine actions to fight against this problem.

 Mathematical modelling is an extremely useful tool to describe different events \cite{emilio}. It can simulate an activity of a system that reproduces, for example, the dynamics of an infectious disease. Then, it's possible to analyze and interpret the results, looking for ways to control epidemics \cite{UNESP}. An accurate model enable political and sanitary authorities to adopt based decisions.

 In literature, a several epidemic models are currently used. Among of them, compartmental ones stand out, dividing population into classes with common characteristics and establishing their relations\cite{murray}. One of them is the $SIR$ model, which divides the population into three classes and proposes a set of differential equations that define the rate of change of each compartment over time \cite{trends}. This change is directly linked to the parameters that make up the equations, as they represent rates that describe the dynamics of disease spread in the population. The differential equations of the $SIR$ model use these constants to describe how the number of individuals in each compartment (susceptible, infected and removed) varies over time. However, for long periods, assigning a single value to these parameters may not be effective. Changes in population behavior, government measures and even the nature of the disease can alter during this period. Thus, the model's ability to reproduce the disease dynamics is compromised.

Based on this, a different approach to the $SIR$ model is proposed in this study. It is assumed that the parameter representing the infection rate varies over time to simulate the behavior of the coronavirus in the state of Rio Grande do Sul for 770 days. Thus, for each day, a value for this parameter was calculated, changing the model from a fixed value to a vector of values representing the infection rates on each of the considered days. To identify these values, data provided by the Rio Grande do Sul Health Department containing the number of infected individuals were used. An optimization algorithm was then applied to minimize the squared difference between the total infected reported by the data and that simulated by the model. However, it was found that the large number of elements in the vector posed difficulties in this identification. Therefore, a strategy was devised to gradually increase the number of entries in the vector, allowing the algorithm to converge to the desired solution. Based on the obtained results, the real behavior of the pandemic in the state of Rio Grande do Sul was compared with that simulated by the identified model.

After identifying the model and its ability to describe the pandemic's behavior over a long period, its validity in predicting disease activity in future periods was also assessed. For this purpose, the effectiveness of the model in predicting disease behavior in scenarios 7, 14, 30, 60, and 180 days ahead was evaluated. To achieve this, the percentage error between the accumulated number of infected individuals in the period and the simulated value from the model identification was calculated.

 This article is disposed as follows: In the next section, a simplified $SIR$ compartmental model is presented, defining its configuration in differential equations system and explaining its dynamics. The simulation of this model is also displayed. Then, in Section 3, the identification of this model based on data provided by state government will be shown. After, in Section 4, the validity of the model to predict the pandemic behavior in future scenarios will be verified. Finally, in Section 5, discussions about the results are made, with the appropriate conclusions.

 \section{$SIR$ Model}

 Mathematical models are important tools to describe and control infectious diseases. In literature, the most used are the compartmental ones, that separate population into classes \cite{UNESP}. These compartments represent the individuals involved in the process and the state which they find themselves.

 The choice of the model should take into account disease properties, such ashow it spreads in the population \cite{UNESP}. 
 A model with lots of compartments could potentially describe better a disease, but usually results in poor predictions due to the complexity of model and difficulty to estimate correctly the large number of parameters. 
 Usually reduced models with two to four compartments, and few parameters, result in better predictions and are preferred to describe infectious diseases.
 
 The $SIR$ model was developed by Kermack and McKendrick \cite{Fred:2010} and propose a simple but efficient configuration to represent diseases with community transmission, those whose origin of infection is undetermined because the transmitting agent circulates throughout the population. The model suggests to express the system behavior by three groups:
 \begin{itemize}
    \item Susceptible (S): Population that can be infected if they have any contact with the disease;
    \item Infected (I): Individuals who are sick and are capable of infecting susceptible;
    \item Removed (R): People who are no longer infected and can't infect any individual anymore. This group is composed by recovered or dead. 
 \end{itemize}

 In that perspective, total population $N$ is considered constant and given by the sum $S(t)+I(t)+R(t)=N$. The dynamics implies that the susceptible people are subject to infection when have contact with an individual from the infected class. Then, they become part of the infected group for a while, when finally don't carry the disease anymore (cure or death) and become part of removed group. This process can be schematized as follows:
 \begin{eqnarray}
S \longrightarrow I \longrightarrow R
\end{eqnarray}

A configuration for a $SIR$ model is expressed by a set of ordinary differential equations:

\begin{eqnarray}
\begin{cases}
\dfrac{dS(t)}{dt} = - \dfrac{\beta I(t)S(t)}{N}\\[.5cm]
\dfrac{dI(t)}{dt} =  \dfrac{\beta I(t)S(t)}{N} - \gamma I(t)\\[.5cm]
\dfrac{dR(t)}{dt} = \gamma I (t)
\end{cases}
\label{eq:sir}
\end{eqnarray}

Notice that in the first equation, which denotes the variation rate of susceptible, contamination of individuals is given by the interaction between $S$ and $I$ classes. 
It is assumed that each individual in the susceptible group has, on average, $e$ encounters with other people every day, and that each encounter with a person in the infected group has a probability $p$ of infecting the person.
This infection rate is denoted by the parameter $\beta=e \cdot r$ and varies with the development of the epidemic, because people change their behavior towards the disease and the disease change over time.
At the same rate that susceptible population leaves the $S$ class, it enters in the $I$ class. Similarly, a portion of the infected population leaves this class, after a mean period $\frac{1}{\gamma}$, and becomes part of the removed group $R$.

%Both infection rate $\beta$ and the removal rate $\gamma$ are proportional constants and, consequently, have values between $0$ and $1$. %However, it's possible to consider this constants varying on time. For example, if we simulate the epidemic behavior for a long time, assigning a single value to the infection rate could be a obstacle to get an accurate result. Therefore, it's interesting that this parameters have different values each month, week or even each day.

\subsection{Analytical Solution}

One of the goals of using mathematical models for epidemic situations is to determine how the disease is spreading. For this purpose, it is interesting to observe the system's behavior over time, motivating oneself to check when the spread will begin to decline.

The $SIR$ model does not have an analytical solution, but one can attempt to understand the evolution of an epidemic through this model when the disease propagation is in its early stages. If, at the beginning of the process ($t=0$), the number of susceptible individuals is almost the entire population ($S \approx N$), and taking $I(0)=I_0>0$, then in the second equation:
\begin{eqnarray}
\dfrac{dI(t)}{dt} = \beta I(t) - \gamma I(t) = I(t)(\beta - \gamma),
\end{eqnarray}
whose solution can be easily found using the method of separation of variables, resulting in:
\begin{eqnarray}
I(t) &=& I_0e^{(\beta - \gamma)t} = I_0e^{\gamma t(\frac{\beta}{\gamma} - 1)}
\end{eqnarray}
Given that $R_0 = \frac{\beta}{\gamma}$, it follows that:
\begin{itemize}
    \item Se $R_0>1$, $I(t) \rightarrow \infty$, that is, epidemic continues to grow;
    \item Se $R_0<1$, $I(t) \rightarrow 0$, that is, the infection tends to end over time.
\end{itemize}

The constant $R_0$ is called the infection reproduction rate.

%and the average infection period is given by $\frac{1}{\gamma}$.

%However, it's possible to consider this constants varying on time. For example, if we simulate the epidemic behavior for a long time, assigning a single value to the infection rate could be a obstacle to get an accurate result. Therefore, it's interesting that this parameters have different values each month, week or even each day.

\subsection{Time-varying infection rate $SIR$ model}

The infection rate of the conventional $SIR$ model represents both: the probability of a susceptible person contracting the disease when coming into contact with an infected individual and the mean number of encounters an susceptible person has with other people every day.  This rate is crucial for understanding how a disease spreads over time. However, assigning a fixed value to this parameter for an extended period can lead to inaccuracies when attempting to simulate the behavior of a disease. This can occur because the model depends on the interaction between susceptible and infected classes. Such interaction may change if a long period of time is considered. Individuals may alter their behavior as the disease spreads, changing the chance of infection. There's also the possibility that authorities may take containment measures, preventing greater contact between individuals in these two classes. Not to mention, the disease itself may mutate, becoming more or less contagious, thereby increasing or decreasing the likelihood of someone contracting it.

From this, an alternative $SIR$ model can be proposed, with a time-varying infection rate. This allows defining how the disease is spreading on a daily, weekly, or monthly basis. To achieve this, the parameter $\beta$ is modified so that instead of receiving a fixed value, it is function over time. Thus, the model is configured as follows:

\begin{eqnarray}
\begin{cases}
\dfrac{dS(t)}{dt} = - \dfrac{\beta (t)I(t)S(t)}{N}\\[.5cm]
\dfrac{dI(t)}{dt} =  \dfrac{\beta(t) I(t)S(t)}{N} - \gamma I(t)\\[.5cm]
\dfrac{dR(t)}{dt} = \gamma I (t),
\end{cases}
\label{eq:sirVaria}
\end{eqnarray}
where $\beta (t)$ is the time-varying infection rate at time $t$.

The cumulative number of infected individuals, is defined by:
\begin{eqnarray}
\hat{Y}(t,\beta(t)) = \int_{0}^{t} \dfrac{\beta(\tau) I(\tau)S(\tau)}{N} ~d\tau = N-S(t)
\end{eqnarray}
where we have explicitly defined $\hat{Y}$ as a function of time $t$ and the time-varying infection rate $\beta(t)$.

\subsection{Simulation of the Model}

Since the model does not have an analytical solution, a computational method can be used to  obtain the numerical solution of the system of differential equations that characterizes the model $SIR$. 
The one chosen in this work was the fourth-order Runge-Kutta \cite{runge}, which solves a problem of the type $y'(t)=f(t,x)$. The sampling period used was fixed, simulating the system's behavior on a daily basis to align with the way the data was grouped. 
Additionally, since the average infection period is given by $1/\gamma$, as seen earlier, this constant can be fixed, knowing that, on average, this period is $10$ days \cite{OMS}. Thus, it was assumed that $\gamma = 0.1$, and, consequently, the simulated $SIR$ model is given by:
\begin{eqnarray} \label{eq:sirF}
\begin{cases}
\dfrac{dS(t)}{dt} = - \dfrac{\beta (t)I(t)S(t)}{N}\\[.5cm]
\dfrac{dI(t)}{dt} =  \dfrac{\beta(t) I(t)S(t)}{N} - 0.1 I(t)\\[.5cm]
\dfrac{dR(t)}{dt} = \gamma 0.1 (t).
\end{cases}\end{eqnarray}

\begin{figure}[h]
    \centering
    \includegraphics[scale=0.75]{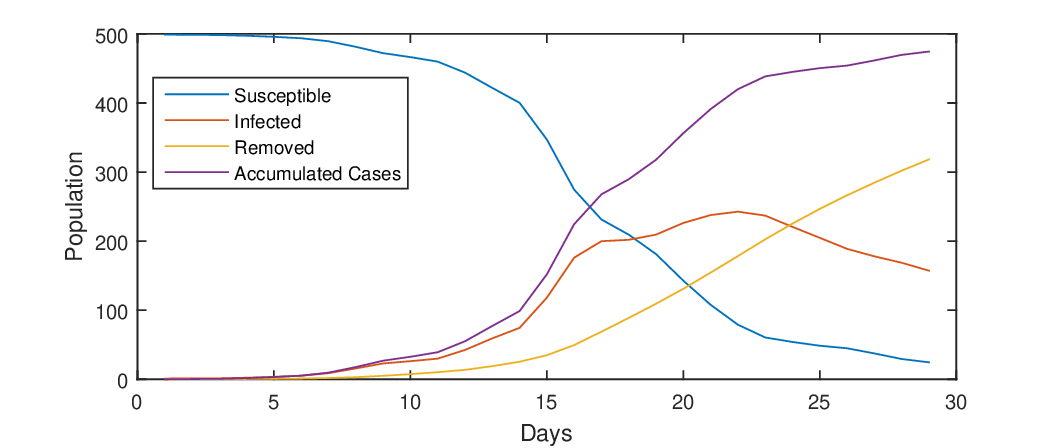}
    \caption{Simulation of the time-varying $SIR$ model}
    \label{fig:simulamod}
\end{figure}

To illustrate the simulated model, the algorithm was executed in a hypothetical case for a population of $N=500$ over 1 month, with randomly varying daily rates. The result can be observed in the graph in Figure \ref{fig:simulamod}.
%%%%%%%%%%%%%%%%%%%%%%%%%%%%%%%%%

\section{Identification of time-varying parameter $\beta(t)$}

Parameter identification involves, based on collected data about the studied phenomenon, determining the values of a parameter in order to approximate the model's behavior to the studied system \cite{bernard}. To achieve this, an optimization algorithm is applied. %citar aqui

The method of least squares was employed in this study. This technique aims to estimates the parameters  that minimizes the sum of a squared function $F$ \cite{wolberg}. In the case where the objective is to fit a mathematical model based on collected data, the function $F$ represents the error between the solution computed by the model and the data that represents the reality. Thus, let $Y(t)$ be the total accumulated infected at the day $t$ and $\hat{Y}(t)$ be the simulated output of the model at the day $t$, the function to be optimized is defined as:
\begin{eqnarray}
J(\beta(t))=\sum_{t=1}^N (Y(t) - \hat{Y}(t,\beta(t)))^2
\end{eqnarray}
where $N$ is the number of days of the collect data.

We can then propose the following optimization problem
\begin{eqnarray}
    \widehat{\beta(t)} = \underset{\beta(t)}{arg ~ min}  ~ J(\beta(t))
\end{eqnarray}

We have also defined that $\beta(t)$ is a piecewise constant function such that it varies from day to day, but it is constant during one day. 
In order to simplify the notation, we have grouped all the parameters (one for each day) as a vector
\begin{eqnarray}
B = \begin{array}{ccccc}[\beta(1) & \beta(2) & \cdots & \beta(N-1) & \beta(N) ]\end{array}
\end{eqnarray}
such that the optimization problems becomes finite:
\begin{eqnarray}
    \widehat{B} = \underset{B}{arg ~ min}  ~ J(B)
\end{eqnarray}

We have used the {\it MATLAB} software to solve this optimization problem with the function {\it fmincon}.

\subsection{Data}
To perform the identification of the model parameters, data provided by the Health Department of the State of Rio Grande do Sul were used, obtained through its website \cite{SecRS}. These data are categorized for each infected person, providing the case registration date and additional information such as gender, age, etc. The considered period spanned 770 days (or 110 weeks), starting from the first registered case.

\begin{figure}[h]
 \centering
     \includegraphics[scale=0.7]{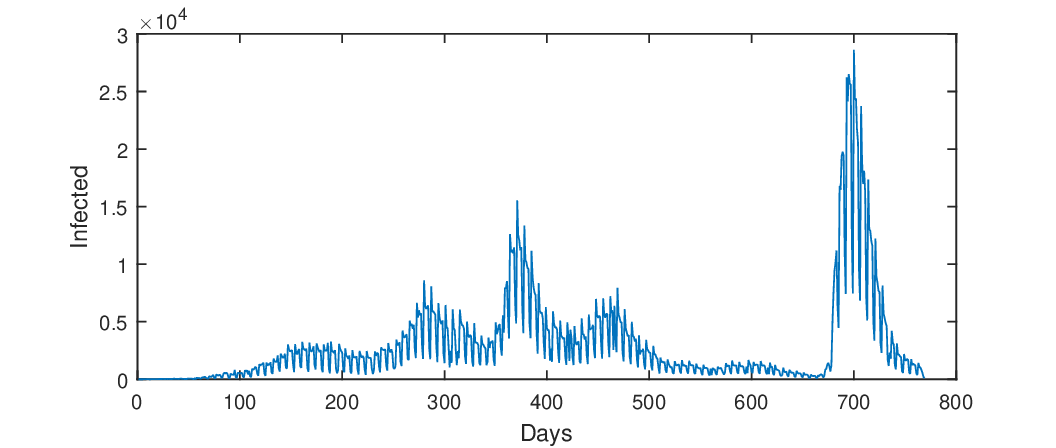}
     \caption{Number of daily cases in RS state}
     \label{fig:nrocasos}
 \end{figure}

\begin{figure}[h]
     \centering
     \includegraphics[scale=.7]{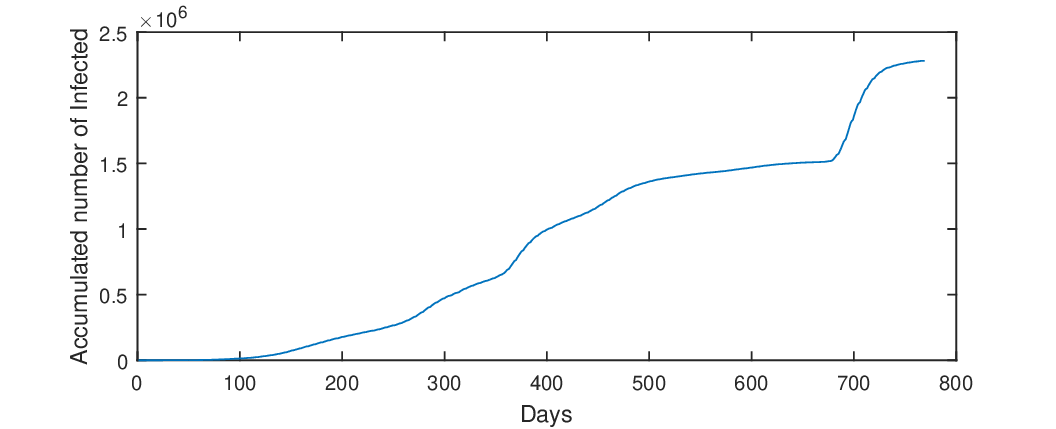}
     \caption{Accumulated number of cases in RS state}
     \label{fig:AcumuladoRS}
 \end{figure}

Based on these records, the software {\it MATLAB} was utilized to create graphs illustrating the number of detected cases each day, as well as the cumulative number of infected individuals over this time period. This allowed the observation of the pandemic's progression in the state, analyzing critical periods, growth, and decline in the number of cases, among other considerations. Such behaviors can be observed in the Figures \ref{fig:nrocasos} e \ref{fig:AcumuladoRS}. From there, it is possible to perform the model identification, as seen earlier, and the results are displayed in the next subsection.

%To achieve the goal of this study, it is necessary to determine the values of the infection rate $\beta(t)$ over a specific period in the State of Rio Grande do Sul. For this purpose, data provided by the Health Department of the State Government were used, which include information about the infected population. The cumulative number of people who contracted the disease is the output of the model, denoted by $Y(t)$ and expressed as:
%\begin{eqnarray}
%Y(t) = \int \dfrac{\beta(t) I(t)S(t)}{N}~dt
%\end{eqnarray}

\subsection{Optimization Algorithm}

The goal of this identification is to find a vector $B$ of $770$ elements, which represents the infection rate corresponding to each day. For instance, the value at position 20 in the vector refers to the infection rate on the 20th day of the pandemic in the state. To determine these values, the cumulative infected data mentioned in the previous subsection were used, and the optimization method was applied to align the simulated graph with the real behavior observed in the Figure \ref{fig:AcumuladoRS}.

Based on the model (\ref{eq:sir}), initial conditions were established for each state variable of the system, and then it was simulated using the Runge-Kutta method. The initial values are:
 \begin{eqnarray}
 \left[\begin{array}{c} S(0) \\ I(0) \\ R(0) \end{array}\right] &=& \left[\begin{array}{c} P-1 \\ 1 \\ 0 \end{array}\right] ,
 \end{eqnarray}
 where $P$ is the approximate total population of Rio Grande do Sul, estimated at 11.3 million.

To optimize the parameter vector $B$, initial values were also assigned to its components, so that the curve of the simulated graph would approximate the real behavior indicated by the data. Since this vector has $770$ elements, calculating all at once results in poor estimates since the numerical solver is trapped in local solutions of the nonconvex optimizaiton problem. Thus, the use of a conventional algorithm becomes ineffective.

To overcome this obstacle, an iterative algorithm was created following these steps: first, it was assumed that the vector $B$ had only one element, meaning the algorithm would calculate only one value to approximate the model's behavior to reality, as if it were a single rate for the entire period. Thus, $\beta$ was assumed to be an initial vector of $[1]$, and the optimization algorithm was executed. Obviously, the error produced when comparing the graphs was large. So, the second step was to duplicate the number of parameters and assign to them the values found by the algorithm in the first optimization. Assuming that the vector $B$ now consisted of 2 values, the first representing the first $512$ days and the second representing the others. The optimization was performed again with the two initial values being (both) those found in the first optimization. In this way, the resulting error was still significant, but the result showed a better fit compared to the previous one. Thus, with each simulation, the number of elements in the vector was doubled, and it was found that the precision began to improve. The table \ref{tab:ident} shows the strategy.

\begin{table}[h]
\begin{tabular}{|c|c|c|c|}
\hline
Iteration &Number of elements & initial vector                                  & final vector                                                  \\ \hline
$1$ &$1$             & $1$                                            & $v_1$                                                        \\ \hline
$2$ &$2$             & $[v_1, v_1]$                                   & $[v_2, v_3]$                                                 \\ \hline
$3$ &$4$             & $[v_2, v_2, v_3, v_3]$                         & $[v_4, v_5, v_6, v_7]$                                       \\ \hline
$4$ &$8$             & $[v_4, v_4, v_5, v_5, v_6, v_6, v_7, v_7]$     & $[v_8, v_9, v_{10}, v_{11}, v_{12}, v_{13}, v_{14}, v_{15}]$ \\ \hline
$$ &$\vdots$        & $\vdots$                                       & $\vdots$                                                     \\ \hline
$10$ &$512$           & $[v_{256}, v_{256}, \cdots, v_{511}, v_{511}]$ & $[v_{512}, \cdots, v_{1023}]$                                 \\ \hline
\end{tabular}
\caption{Iterative strategy to solve the optimization problem}
\label{tab:ident}
\end{table}

\begin{figure}[h]
    \centering
\includegraphics[scale=.7]{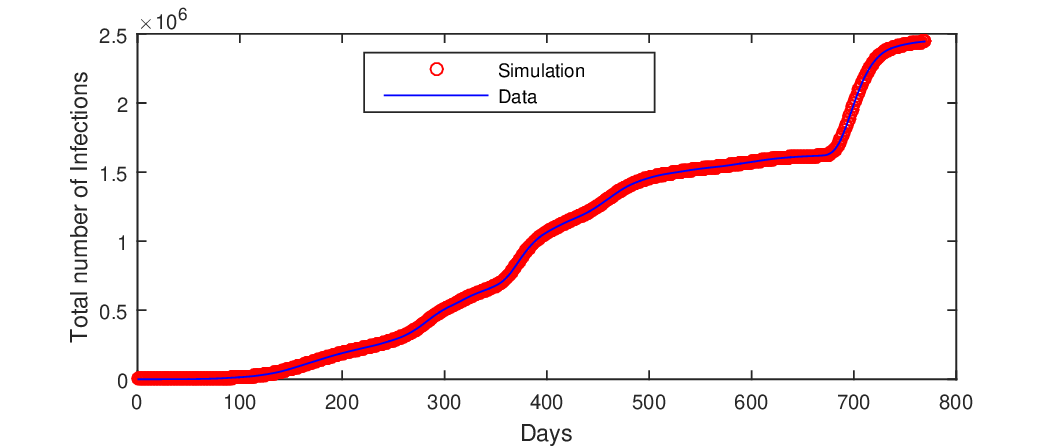}
    \caption{Real Behavior x Simulated}
    \label{fig:identificada}
\end{figure}

\begin{figure}[h]
    \centering
    \includegraphics[scale=0.7]{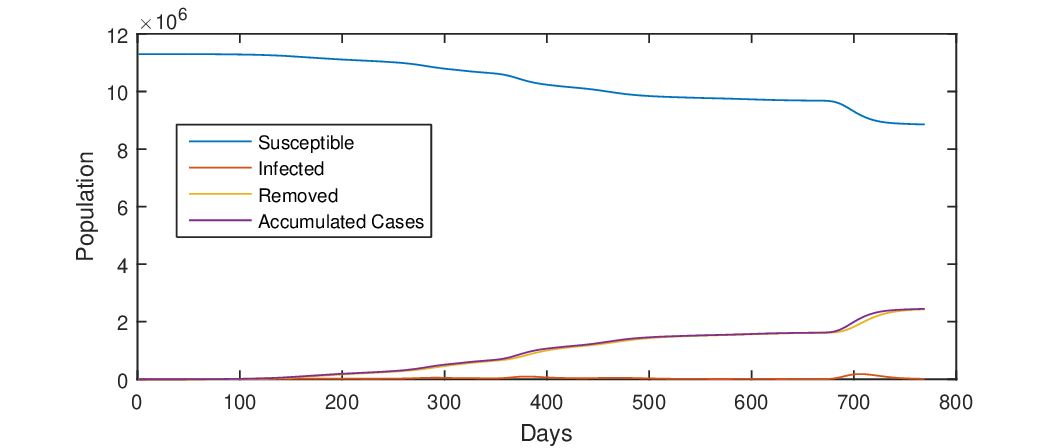}
    \caption{Simulation with identified infection rates}
    \label{fig:simulareal}
\end{figure}

In the last step, after obtaining a final vector with $1024$ elements, values from the 771st position to the last were discarded, and then one rate was optimized for each day. The purpose of this process was to approximate the behavior by gradually adjusting the vector, adding elements. In the end, the vector $B$ with the identified rates was obtained. As a result, the model achieved the intended accuracy, as seen in Figure \ref{fig:identificada}, where the graphs are compared.
Simulating the model based on the identified rates, the system's behavior is observed in the Figure \ref{fig:simulareal}.

With the use of the data, it was possible to determine the infection rates for each day of the pandemic, allowing the behavior and evolution of infections to be observed with good precision. However, this technique reveals what happened in the past. Is it possible to understand how this behavior will unfold in the future? Motivated by this question, an investigation was carried out to assess how effectively this model is to to predict the disease activity in the days following the known data.

\section{Prediction of a $SIR$ Model}

In this work, the aim is to determine how accurate the prediction of the $SIR$ model can be based on the identification of its infection rates. Thus, given the behavior of the cumulative infected individuals in the previous section, the accuracy of the model in forecasting its activity over different time intervals was assessed.

\subsection{Results}

The process for evaluating the model at future time intervals was carried out as follows: it was assumed that the pandemic was already on its 100th day and that the infection rate for each $100$ days of the pandemic was estimated.
In order to estimate the following days of the pandemic, it was assumed that from that day on, the value $\beta(t)$ would be constant, for $t>100$. Since it was known that the infection rate would change in the future, it was certain the model would be accurate in the following days but it would diverge in the future. 
The goal is then determine the accuracy of the model in predicting the cumulative behaviour of the pandemic over the next $7$, $14$, $30$, $60$, and $180$ days. The percentage of error was computed as follow
\begin{eqnarray*}
    e_T(t) = \dfrac{Y(t+T) - \hat{Y}(t+T)}{Y(t+T)}
\end{eqnarray*}
where $T=\{7, 14, 30, 60, 180\}$, such that five metrics were evaluated.

Subsequently, the same procedure was repeated starting from the 101st day, analyzing the simulation for the same subsequent periods (7, 14, 30, 60, and 180 days). This process was then repeated starting from the 102nd day and continued successively. The analysis was conducted by examining the percentage error between the prediction attempt and the previously identified behavior, as given by:
For each time interval, the average percentage error and standard deviation were also computed.

\subsubsection{Prediction for $T=7$ days}

As reported earlier, the percentage error was examined between the identified model's cumulative number of infected individuals and the data, 7 days ahead. These errors were tallied and distributed in a histogram. Additionally, the average percentage error of all simulations, along with its standard deviation, was calculated to subsequently compare the histogram with the Gaussian curve.
\begin{figure}[h]
    \centering
    \includegraphics[scale=0.75]{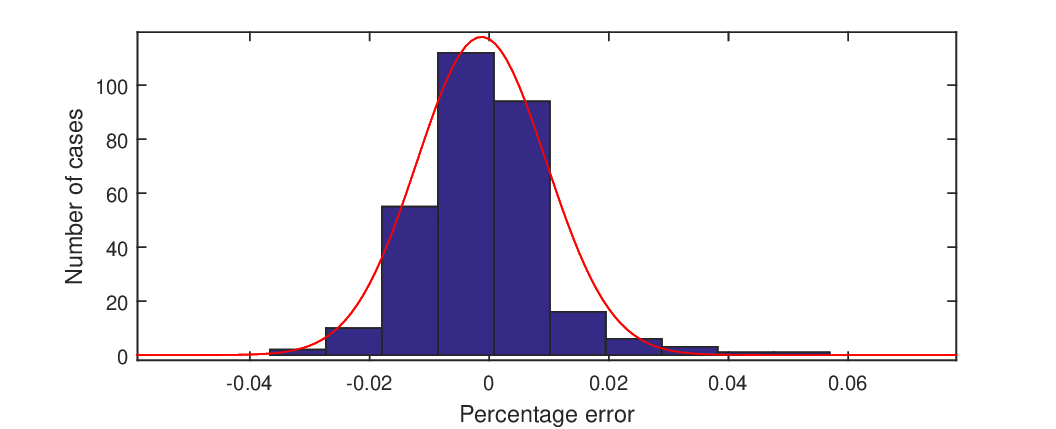}
    \caption{Percentual error 7 days in future}
    \label{fig:7d}
\end{figure}

As can be seen, the prediction of infection rates $7$ days ahead exhibited a behavior close to that of the Gaussian curve, with a calculated mean of the percentage errors at $0.13\%$ and a standard deviation of $1.08\%$. As noted, the majority of cases are within one standard deviation of the mean, indicating the model's good accuracy in forecasting the pandemic's behavior one week into the future.

\subsubsection{Prediction for 14 days}

\begin{figure}[h]
    \centering
    \includegraphics[scale=.75]{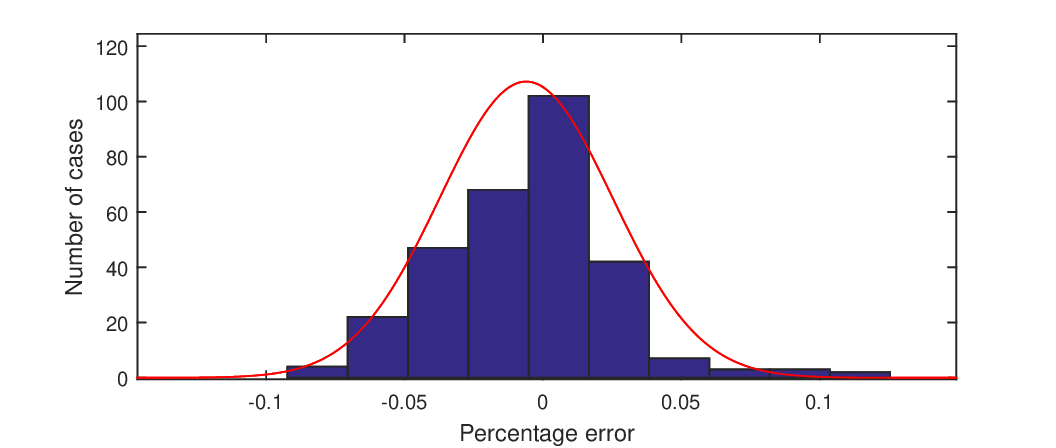}
    \caption{Percentage error 14 days in future}
    \label{fig:14d}
\end{figure}

After verifying the effectiveness of the model in predicting the evolution of the pandemic one week ahead, we sought to analyze its performance two weeks into the future. The graph in the Figure  \ref{fig:14d} illustrates the percentage errors of the simulations.

Knowing that the mean of the percentage errors was $0.6\%$ and the standard deviation was $3.13\%$, it is possible to observe that, once again, the histogram's behavior resembled a Gaussian curve. However, a higher number of simulations with errors exceeding one standard deviation from the mean were noticed, indicating that the model's accuracy was not as good as in the previous situation.

\subsubsection{Prediction for 30 days}

\begin{figure}[h]
     \centering
     \includegraphics[scale=.75]{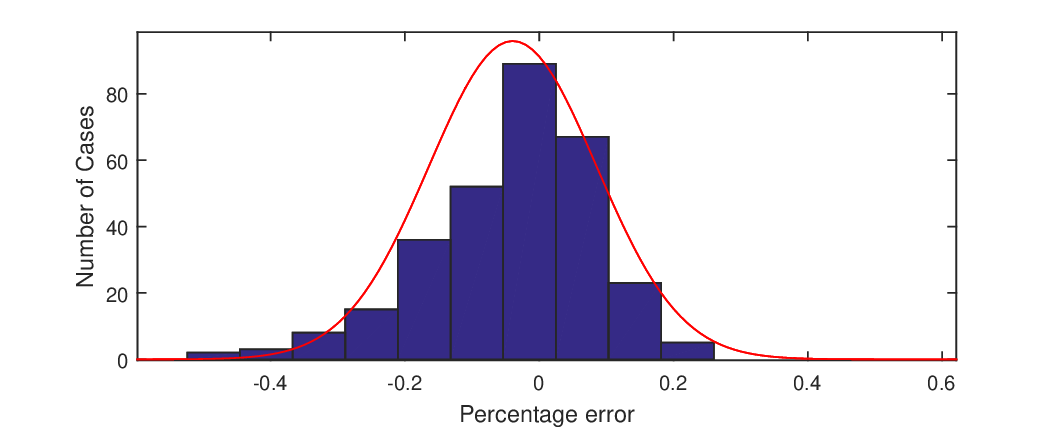}
     \caption{Percentual error 30 days in future}
     \label{fig:30d}
 \end{figure}

Here, the percentage error produced by the model in attempting to simulate the pandemic activity one month into the future was examined. This resulted in an average percentage error of $3.95\%$, with a standard deviation of $12.48\%$. Thus, it can be observed that, despite the average not being high, there were many cases far from the mean, and the distribution of cases did not resemble that of the Gaussian curve, as can be seen in the Figure \ref{fig:30d}.
 Thus, it is understood that the model did not exhibit fidelity to the actual behavior of the pandemic when attempting to predict it 30 days ahead.

 \subsubsection{Prediction for 60 days}

 In this subsection, the results of the model simulation in attempting to predict the evolution of the cumulative number of infected individuals in the 60 days following the already known period are presented. It was determined that the average percentage error of the simulated behavior, in comparison to the actual data, was approximately $22.97\%$, indicating a low level of precision. Additionally, the standard deviation found was $53.32\%$.
 \begin{figure}[h]
     \centering
     \includegraphics[scale=.75]{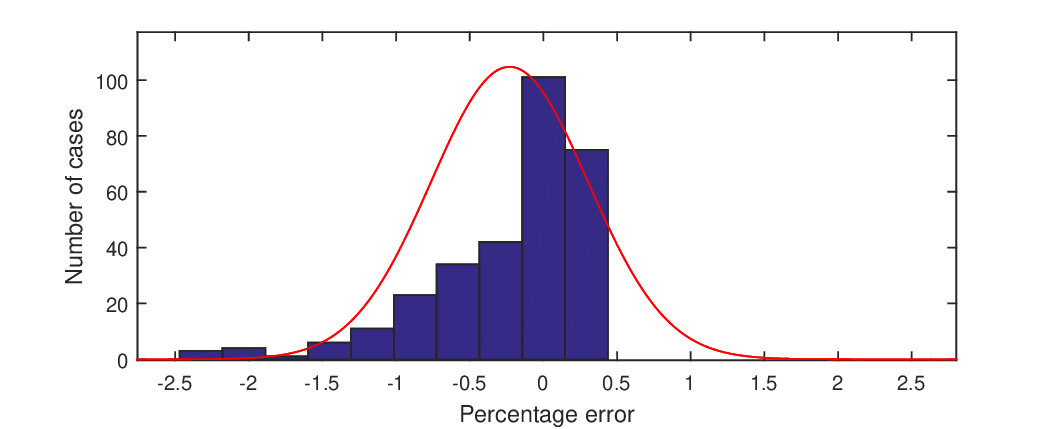}
     \caption{Percentual error 60 days in future}
     \label{fig:60d}
 \end{figure}
 
 When comparing the quantity of cases distributed in the histogram with its Gaussian curve, it is observed that there is no proximity between the graphs, as illustrated in Figure \ref{fig:60d} Therefore, it is evident that the model is not reliable in predicting how the pandemic will unfold two months into the future.

 \subsubsection{Prediction for 180 days}

\begin{figure}[h]
     \centering
     \includegraphics[scale=.75]{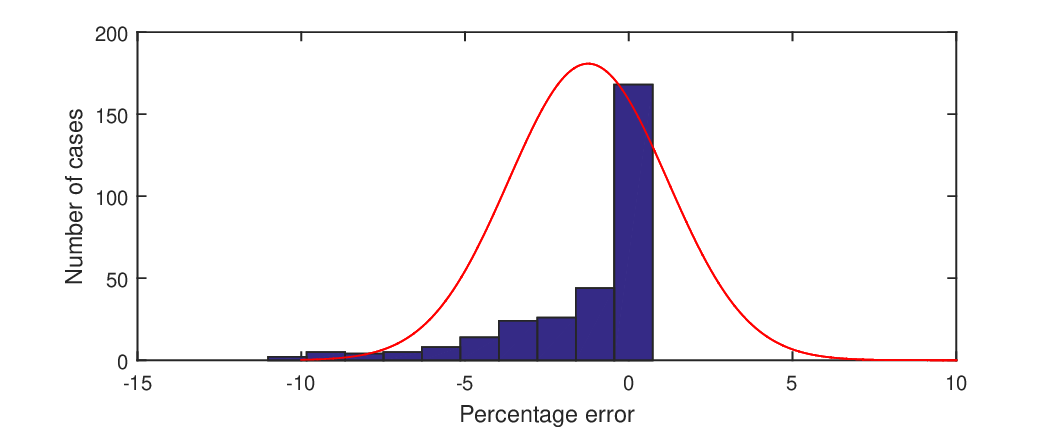}
     \caption{Percentual error 180 days in future}
     \label{fig:180d}
 \end{figure}

 After realizing that the model did not exhibit accuracy in simulating the behavior of the pandemic for 1 or 2 months into the future, a new test was conducted to assess its activity 6 months ahead. As expected, it was observed that the percentage error was significantly high in practically all cases, as can be seen in Figure \ref{fig:180d}, depicting the distribution of cases in this scenario alongside the corresponding Gaussian curve.

 For this case, the average percentage error was $123\%$, with a standard deviation of $242\%$, confirming the impracticality of using the model to predict the disease's behavior during this time period.

 \section{Conclusions}

 Epidemics are phenomena that frequently arise in populations. Understanding the dynamics of diseases that manifest collectively is crucial for implementing prevention and combat measures \cite{ecology}. In this regard, mathematical models emerge as useful tools for this purpose.

In this study, the use of a simplified compartmental model $SIR$ was proposed to describe the behavior of the coronavirus pandemic. However, the conventional system of differential equations in this model suggests constants with fixed values to represent the infection rate, which can lead to inaccuracies in simulating long periods of time. Because of this, a time-varying infection rate was adopted to reflect the daily activity of the disease. Thus, for each day, a value for this constant was identified. 

In this identification, data on the population infected by Covid-19 over a period of 770 days in the state of Rio Grande do Sul were used. An optimization algorithm, based on the method of least squares, was executed to determine the infection rates on each day of analysis. Since the daily simulation required a vector with many parameters and the optimization problem was non-convex, it was necessary to develop an iterative algorithm that doubled the number of elements in the vector to be identified after each iteration. 

This approach allowed for adjusting the model to the data and obtaining good accuracy in representing the accumulated number of infected individuals, as observed in the graphs comparing the simulation with the real behavior generated by the data.  The identified model has a very good accuracy and represents very well the development of the pandemic, which means that it is complex enough to represent the evolution of the disease, and that models with more compartments are not necessary.

A study was also conducted on the model's validity in predicting the disease's activity in future scenarios. Thus, assuming that the pandemic's behavior in RS was unknown from a certain day onwards, the model was used to simulate its development for 7, 14, 30, 60, and 180 days into the future.
It was also found that the model can reasonably accurately describe pandemic scenarios one or two weeks into the future. To do this, the percentage error between the cumulative number of infected individuals in the period and the estimate by the model was calculated. By simulating the disease activity 7 days ahead, an error of less than $0.13\%$ was obtained, indicating good predictive capability. For 14 days, the generated percentage error was less than $0.6\%$, signaling that the model can also effectively predict the behavior of the pandemic in this period. However, for time intervals greater than 14 days, the use of the model for prediction is not recommended, as the generated percentage errors were large.

The main source of error in predicting the future was the choice of constant value for all days ahead. It was observed in the identification process that the infection rate varies a lot from one day to day. A future research topic is to choose other types of functions for the upcoming days.

\bibliographystyle{ieeetr}
\bibliography{TCAM_modelo}
	
\end{document}